# Guided resonances in photonic crystals with point-defected aperiodically-ordered supercells

**Ilaria Gallina,[1] Marco Pisco,[2] Armando Ricciardi,[3] Stefania Campopiano,[3] Giuseppe Castaldi,[1] Andrea Cusano,[2] and Vincenzo Galdi [1,*]**

[1] *Waves Group, Dept. of Engineering, University of Sannio, Corso Garibaldi 107, I-82100 Benevento, Italy*
[2] *Optoelectronic Division, Dept. of Engineering, University of Sannio, Corso Garibaldi 107, I-82100 Benevento, Italy*
[3] *Dept. for Technologies, University of Naples "Parthenope," Centro Direzionale, Isola C4, I-80143, Naples, Italy*
*\* vgaldi@unisannio.it*

**Abstract:** In this paper, we study the excitation of *guided resonances* (GRs) in photonic-crystal slabs based on point-defected *aperiodically-ordered* supercells. With specific reference to perforated-slab structures and the Ammann-Beenker octagonal lattice geometry, we carry out full-wave numerical studies of the plane-wave responses and of the underlying modal structures, which illustrate the representative effects induced by the introduction of symmetry-preserving and symmetry-breaking defects. Our results demonstrate that breaking the supercell mirror symmetries via the judicious introduction of point-defects enables for the excitation of otherwise uncoupled GRs, with control on the symmetry properties of their field distributions, thereby constituting an attractive alternative to those GR-engineering approaches based on the asymmetrization of the hole shape. In this framework, aperiodically-ordered supercells seem to be inherently suited, in view of the variety of inequivalent defect sites that they can offer.

**OCIS codes:** (160.5298) Photonic crystals; (260.5740) Resonance.

## 1. Introduction

Photonic crystals (PCs) are artificial periodic (metallic or dielectric) structures known as the electromagnetic (EM) counterpart of semiconductors, in view of their potential capability of controlling to a great extent the EM wave propagation [1]. For optical engineering applications, two-dimensional (2-D) PC slab configurations are typically based on periodic arrangements of air holes in a host medium. Such structures are capable of exhibiting anomalous interactions with the EM fields of great interest for both basic and applied science. For instance, typical effects observable in connection with *in-plane* propagation include bandgaps [2], "negative" refraction [3], and subwavelength focusing ("superlensing") [4], whereas, in connection with *out-of-plane* propagation, it should be mentioned the possibility of exciting "guided resonances" (GRs) [5].

During the past decade, following up on some recent discoveries in solid-state physics [6,7], and based on the theory of "aperiodic tilings" [8], the study of the above effects has been extended to *aperiodically-ordered* structures, called "photonic quasicrystals" (PQCs), confirming the possibility of obtaining similar properties as those exhibited by PCs, with significant potential improvements attainable via a judicious exploitation of the additional degrees of freedom inherently available in aperiodic geometries (see, e.g., [9-11] for recent reviews of the subject).

In an ongoing series of recent investigations [12,13], we have explored the mechanisms underlying the excitation of GRs in PQC slabs. In particular, in [12], we showed the possibility of exciting GRs in PQC slabs based on a *quasiperiodic* (Ammann-Beenker, octagonal) defect-free lattice exhibiting eight-fold symmetry [8]. Subsequently, in [13], we explored the dependence of such GRs on the main structural parameters (air/dielectric fraction, slab refractive index and thickness). In agreement with what observed in the periodic PC case [5], and against the known evidences that GR may be destroyed by lattice *disorder* [14], the above results indicate can they also be induced by *aperiodic order*, and are still attributable to the coupling of (degenerate, properly-symmetric) *leaky modes* supported by the PQC slab with the continuum of free-space modes. Moreover, as observed in connection with plane-wave-excited periodic PC slabs [15], GRs show up in the transmittance response as sharp, asymmetric Fano-like [5,16,17] resonant line shapes superposed to a smoothly varying background (attributable to Fabry-Perot effects due to an effectively-homogeneous dielectric slab), and turn out to exhibit similar dependences on the slab parameters [13,18].

In this paper, we study the excitation of GRs in PC slabs characterized by *point-defected* (in the form of missing holes) *aperiodically-ordered* supercells. Lattice defects in PCs and PQCs have been extensively studied in connection with typical in-plane propagation phenomena such as field localization and waveguiding, and the potential advantages of using aperiodic lattices (in terms, e.g., of richer and more wavelength-selective states) have been pointed out (see, e.g., [19-21] for a sparse sampling). Conversely, the role of defect states in the GR scenario turns out to be still largely unexplored, and yet potentially capable of opening up new perspectives in their control/tailoring. Within this context, previous studies on periodic PCs [22] have focused on control mechanisms based on breaking the mirror symmetries of (all) the air holes, demonstrating the possibility of exciting "uncoupled" GRs with prescribed symmetry properties [23]. Such approaches, while proven to be technologically viable, are inherently limited by the fabrication tolerances. Intuitively, a control mechanism based on the introduction of *localized* lattice defects would appear more versatile and robust, and would therefore constitute an attractive alternative. Although, in principle, such approach may likewise be implemented via point-defected *periodic* supercells, from the in-plane propagation experience, one would expect it to work much more effectively with *aperiodically-ordered* supercells, in view of their inherent richness of inequivalent defect sites. Accordingly, we focus here on the quasiperiodic supercell geometry considered in [12,13], based on the Ammann-Beenker (octagonal) tiling.

The rest of the paper is structured as follows. Section 2 introduces the problem geometry and statement. Section 3 provides some details about the simulation parameters and computational tools utilized throughout. Section 4 is devoted to the illustration and discussion of representative results from full-wave numerical simulations concerning the transmittance response and modal analysis of defect-free and various (symmetry-breaking or symmetry-preserving) defected configurations. Finally, Section 5 addresses some brief conclusions and perspectives.

## 2. Problem geometry and statement

As in [12,13], the PC slab under study is constructed starting from a dielectric slab of thickness $h$ immersed in air, and placing air holes (of radius $r$) at the vertices of a quasiperiodic lattice based on the octagonal Ammann-Beenker tiling made of square and rhombus tiles (see [8,12,24] for details about the generation algorithm). Figure 1(a) shows a 3-D view of a defect-free supercell containing 21 holes (or fractions of them), with indication of the lattice constant $a$ (square/rhombus tile sidelength [8]) and the total sidelength $L = \left(2 + \sqrt{2}\right)a$, from which its mirror symmetries are evident.

As anticipated, our investigation is aimed at exploring the possibility of exciting "uncoupled" GRs with prescribed spatial symmetry properties. Unlike standard approaches based on the asymmetrization of the hole shape [22], our proposed strategy relies on the introduction of localized lattice defects in the form of missing holes, and therefore appears to be potentially less sensitive to fabrication tolerances. Accordingly, from the defect-free configuration in Fig. 1(a), not considering the hole fractions at the boundaries of the supercell, we can generate four inequivalent (for a given polarization of the illuminating field) point-defected configurations, as shown in Figs. 1(b)-1(e). As previously mentioned, one may likewise think of introducing point-defects in *periodic* supercells. However, it is fairly evident that, in such configurations, the only degree of freedom available would be the supercell size since, in view of the underlying periodicity, the local geometry surrounding a defect would be the same, irrespective of its position. Thus, changing the point-defect position would only yield irrelevant spatial translations in the field distribution. Conversely, an aperiodically-ordered supercell like that shown in Fig. 1(a), albeit relatively small, may admit a variety of *inequivalent* (i.e., surrounded by *different* local geometries) defect sites, thereby constituting an interesting prototype configuration inherently suited to explore the proposed symmetry-breaking-based approach to GR tailoring/control.

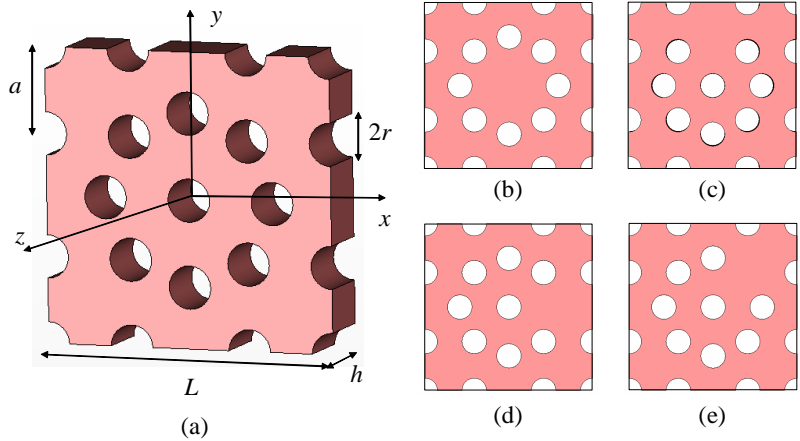

Fig. 1. Problem geometry. (a) 3-D view of the defect-free aperiodically-ordered supercell based on the Ammann-Beenker octagonal tiling, in the associated Cartesian reference system. (b)-(e) 2-D views (*x-y* plane) of the representative point-defected configurations considered.

Our numerical studies are focused on the EM response of the supercells in Fig. 1 (terminated by Bloch-type periodic boundary conditions) for time-harmonic $(\exp(-i\omega t))$ normally-incident plane-wave excitation with *y*-polarized electric field. Moreover, in order to gain a deeper insight into the underlying coupling mechanisms, we also carry out a modal analysis.

It is important to stress that the structures under consideration are *periodic* PCs obtained by replicating an *aperiodic* supercell of relatively small size. Although *short-range* interactions have been observed to play an important role in connection with in-plane propagation/localization phenomena in PQCs (see, e.g., [10,21,25,26]), extension of our results to actual PQC structures is not necessarily meaningful within the point-defect framework. In fact, one would intuitively expect the effect of a simple point-defect to become *progressively weaker* as the supercell size is increased (see also the discussion in Section 4.5 below).

### 3. Simulation parameters and computational tools

In our simulations, we consider a silicon slab (refraction index *n*=3.418) of thickness $h = 0.75a$, and a hole radius $r = 0.25a$ (corresponding to an air/dielectric fraction of about 24% in the defect-free supercell of Fig. 1(a)).

As in [12], we use the commercial software packages CST MICROWAVE STUDIO ® [27] (based on the finite-integration technique) and RSOFT BandSOLVE ™ [28] (based on a 3-D plane-wave expansion method) to study the plane-wave response and the (source-free) modal solutions, respectively.

For both analyses, the computational domains are composed of the supercells in Fig. 1 sandwiched between two air layers of thickness 40*h* along the *z*-axis, and are terminated by Bloch-type periodic boundary conditions in the transverse (*x-y*) plane. At the air-layer terminations along the *z*-axis, periodic boundary conditions are also enforced for the modal analysis [28], while open-boundary (matching) conditions are enforced for the study of the plane-wave response [27]. In order to illustrate the basic, and yet representative, phenomenologies, we focus our study on the normalized frequency range $0.13 \le \nu \le 0.15$, where $\nu = \omega a/(2\pi c)$ (with *c* denoting the speed of light in vacuum). Recalling that the normalized cutoff frequency of the (*p*,*q*)-th order Bloch mode associated with a square lattice of period *L* is given by

$$\nu_{p,q} = \sqrt{\left(\frac{pa}{L}\right)^2 + \left(\frac{qa}{L}\right)^2}, \quad p,q = 0, \pm 1, \pm 2, \ldots, \tag{1}$$

and thus the normalized cutoff frequency of the first higher-order mode is $\nu_{1,0} = a/L \approx 0.293$, it is readily verified that, in the above frequency range, all the higher-order Bloch modes are *evanescent*, and thus the underlying coupling mechanisms are essentially mediated by the normally-incident plane-wave excitation (corresponding to the nondispersive (*p*=0,*q*=0) mode in (1)).

In the plane-wave-response simulations, the computational domain is discretized using an adaptive tetrahedral mesh refinement [27], and the frequency sampling of the transmittance response is also chosen adaptively (up to steps as fine as $\Delta\nu \sim 10^{-5}$, so as to resolve the narrowest resonances). Note that, in view of the *symmetry-breaking* nature of some of the defect configurations, reliance cannot be made on the reduced model (one quarter of the structure placed in a "waveguide simulator" bounded by perfect electric/magnetic walls) exploited in [12,13], and thus the *entire* supercell needs to be studied.

The modal analysis is carried out by studying the band-structures at the Γ point of the irreducible Brilluoin zones associated to the supercells (corresponding to the normally-incident plane-wave excitation of interest). In particular, we consider modal solutions exhibiting both *even* (transverse-electric-like) and *odd* (transverse-magnetic-like) symmetry along the *z*-axis, focusing the attention on the eigen-frequencies (and related eigen-modes) falling within the spectral range of interest. In this framework, it should be noted that the arising solutions include, besides the sought leaky modes, also a number of *radiative* modes from the continuum spectrum that are "artificially" discretized by the assumed periodic boundary conditions along the *z*-axis. These artifacts can easily be identified and discarded by looking at their field distributions along the *z*-axis, which are inherently *extended* (at variance with the leaky modes, which are mostly localized within the slab). Equivalently, one may look at the eigen-frequency sensitivity with respect to changes in the air-layer thickness, identifying the (termination-independent) leaky modes as the *stable* solutions, and the (termination-dependent) artifacts as the *unstable* ones.

## 4. Representative results

### 4.1 Generalities

In what follows, we illustrate the outcomes of our full-wave studies on the five configurations in Fig. 1. Representative results are displayed in Figs. 2-12, in the form of transmittance responses (transmission-coefficient squared-magnitude as a function of the normalized frequency) and electric-field maps (both modal and plane-wave excited solutions). The field maps pertain to the (dominant) *x*- and *y*-components of the electric field, computed at the plane *z*=0; the considerably weaker *z*-components are not shown. For conciseness, the modal field maps are shown only in connection with the excited resonances. However, results from the modal analysis are compactly summarized in Table 1, in the form of mode eigen-frequencies and basic symmetry properties of the electric field *y*-component (directly relevant to the coupling mechanisms). For each configuration, four *z*-even (transverse-electric-like) leaky modes are found within the frequency range of interest, and are labeled by a letter (a-e, tagging the corresponding configuration in Fig. 1) and an integer (1-4, for increasing eigen-frequency ordering). No *z*-odd (transverse-magnetic-like) leaky mode is found within the frequency range of interest. The resonances observed in the transmittance responses can be associated to the leaky modes supported by the structures by matching the spatial field distributions observed in the plane-wave response with those obtained from the modal analysis. Recalling the even symmetries (along both the *x*- and *y*- axes) and the assumed polarization of the plane-wave excitation (*y*-directed electric field), the modal coupling/uncoupling can easily be ascertained by looking at the symmetry properties of the *y*-components of the electric fields (see [12] for more details) in the transverse plane. In particular, the coupling is possible in the presence of even symmetries, whereas it is prevented by the presence of an odd symmetry (along either the *x*- or *y*-axis). The most general case of *non-symmetric* (i.e., neither even nor odd) field distributions can always be viewed as the superposition of even and odd parts, whereby the even part may establish the coupling. Table 1 also reports these associations, with indication of the resonant frequencies extracted from the transmittance spectra via a Fano-type fit (see [5,12,17] for details).

Table 1. Summary of results from the modal analysis and the study of the transmittance responses. The basic properties (normalized eigen-frequencies and relevant symmetries) of the leaky modes supported by the structures are summarized (with reference to the electric-field *y*-component, directly relevant for the coupling), and are matched with the resonances observed in the transmittance responses.

| Configuration | Modal analysis | | | Transmittance response |
|---|---|---|---|---|
| | Mode label | Eigen-frequency | Symmetry (*y*-component) | Resonant frequency |
| (a): defect-free | a1 | 0.146 | *x*- odd, *y*-odd | - |
| | a2 | 0.146 | *x*-even, *y*-even | 0.1454 |
| | a3 | 0.146 | *x*-odd, *y*-even | - |
| | a4 | 0.146 | *x*-odd, *y*-even | - |
| (b): defected, symmetries preserved | b1 | 0.142 | *x*- odd, *y*-odd | - |
| | b2 | 0.142 | *x*-even, *y*-even | 0.1421 |
| | b3 | 0.142 | *x*-odd, *y*-even | - |
| | b4 | 0.142 | *x*-odd, *y*-even | - |
| (c): defected, *y*-symmetry broken | c1 | 0.142 | *x*-even | 0.1411 |
| | c2 | 0.144 | *x*-odd | - |
| | c3 | 0.146 | *x*-odd | - |
| | c4 | 0.148 | *x*-odd | - |
| (d): defected, *x*-symmetry broken | d1 | 0.142 | *y*-odd | - |
| | d2 | 0.144 | *y*-even | 0.1435 |
| | d3 | 0.146 | *y*-even | 0.1453 |
| | d4 | 0.148 | *y*-even | - |
| (e): defected, *x*- and *y*-symmetries broken | e1 | 0.143 | none | 0.1429 |
| | e2 | 0.144 | none | 0.1432 |
| | e3 | 0.146 | none | 0.1453 |
| | e4 | 0.147 | none | 0.1457 |

*4.2 Defect-free configuration*

We start considering the defect-free configuration in Fig. 1(a). From the transmittance response shown in Fig. 2, a Fano-like resonant line shape is clearly observable within the frequency range of interest, with the corresponding (modal and plane-wave-excited) field maps shown in Fig. 3. As expectable in view of the supercell mirror symmetries, from the modal analysis (cf. Table 1), the structure is found to support two pairs (a1-a2 and a3-a4, respectively) of *doubly degenerate* modes (obtainable from each other via interchange of the *x*- and *y*-components and 90°-rotation of the spatial distributions; see also [12] for more details). In particular, three of these modes (a1, a3, and a4) exhibit *odd* symmetries along the

*x*- and/or *y*- axes (which clearly prevent their coupling), while the remaining one (a2, shown in Figs. 3(c) and 3(d)), exhibits the *even* symmetry properties in the *y*-component (along both axes) that render its coupling possible. Indeed, the corresponding eigen-frequency is in good agreement with the resonant frequency extracted from the transmittance response in Fig. 2, and its spatial distribution matches very well the plane-wave-excited one at resonance (see Figs. 3(a) and 3(b)).

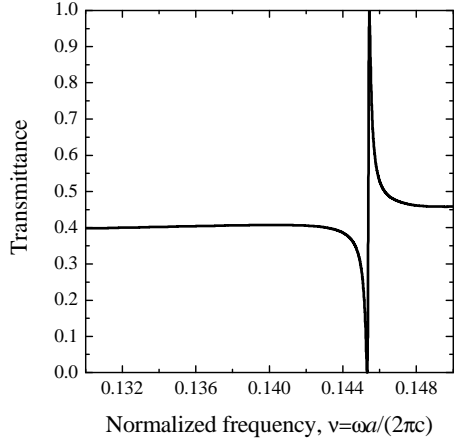

Fig. 2. Transmittance response (for normal plane-wave incidence, with *y*-polarized electric field) pertaining to the defect-free supercell in Fig. 1(a), with $r=0.25a$ and $h=0.75a$.

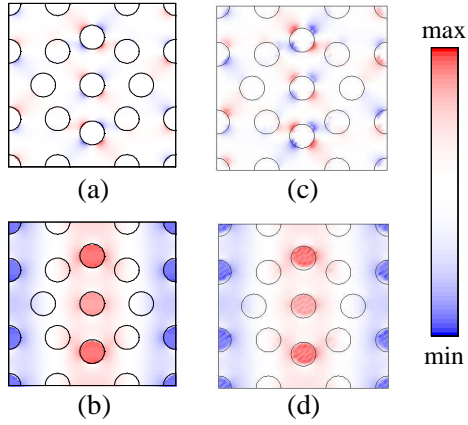

Fig. 3. Parameter configuration as in Fig. 2. (a), (b) Electric field maps (*x*- and *y*-component, respectively) at the resonant frequency $\nu = 0.1454$, and $z=0$. (c), (d) Corresponding modal solution (labeled as a2 in Table 1).

### 4.3 Symmetry-preserving defect

Next, we consider the geometry in Fig. 1(b), featuring a *central* point-defect, which therefore preserves the mirror symmetry properties of the previous defect-free configuration. From the results shown in Figs. 4 and 5, one observes that the introduction of such a defect induces a rather *localized* modification (around the defect site) in the spatial field distribution, as compared with the spatial distribution observed in the defect-free case (cf. Fig. 2), resulting into a moderate down-shift and narrowing of the corresponding resonance in the transmittance response (Fig. 4). Considerations similar to those pertaining to the defect-free case also hold for the modal analysis. Once again, two pairs of doubly degenerate leaky modes are found (b1-b2, and b3-b4, respectively, cf. Table 1), three of which (b1, b3, and b4) are uncoupled in view of their odd symmetries, while the remaining one (b2) exhibits a close match (both in the field distribution and eigen-frequency) with the resonance observed in Fig. 4.

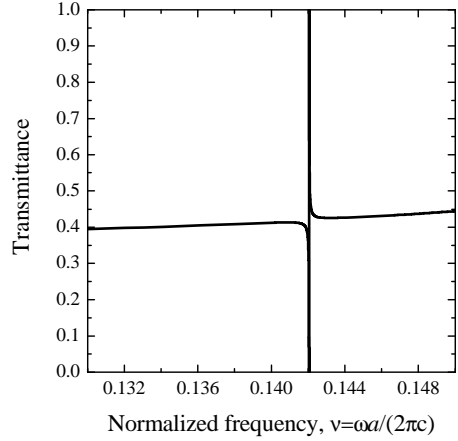

Fig. 4. As in Fig. 2, but pertaining to the point-defected supercell in Fig. 1(b).

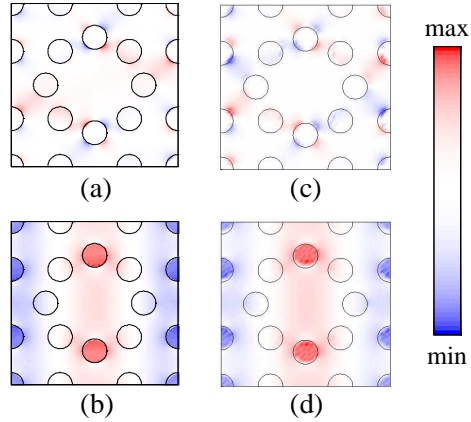

Fig. 5. As in Fig. 3, but pertaining to the point-defected supercell in Fig. 1(b). The resonant frequency is $\nu = 0.1421$, and the modal solution is labeled as b2 in Table 1.

*4.4 Symmetry-breaking defects*

Recalling the findings in [22], much more interesting (in a perspective of exciting "uncoupled" GRs) appear those configurations featuring point-defects that break the supercell mirror symmetries. Basically, one expects the introduction of defects to perturb the spatial distributions of the modal fields supported by the defect-free configuration in Fig. 1(a), so as to destroy the odd symmetries that prevent the modal coupling. However, not all lattice perturbations are equally effective in this respect.

For instance, the configuration in Fig. 1(c) features a point-defect that breaks the mirror symmetry along the *y*-axis, while preserving that along the *x*-axis. This eliminates the previously observed mode degeneracy, resulting into four *distinct* leaky modes supported by the structure, as shown in Table 1. Nevertheless, as can be observed from Fig. 6, there is still only one GR excited. The underlying (un)coupling mechanism can be understood from Table 1, noting that only one mode (c1) exhibits an even symmetry along the *x*-direction (and can therefore couple to the impinging plane wave), while the remaining three (c2, c3, and c4) still exhibit odd symmetries that prevent their coupling. By comparison with the previous two symmetric cases (Figs. 2 and 4), a further down-shift and narrowing of the resonance can be observed, while the effects in the field distribution remain mostly localized around the defect area (cf. Fig. 7). As for the previous configuration, a good agreement is observed (both in the resonance frequency and the field distribution) between the plane-wave response and the corresponding modal solution.

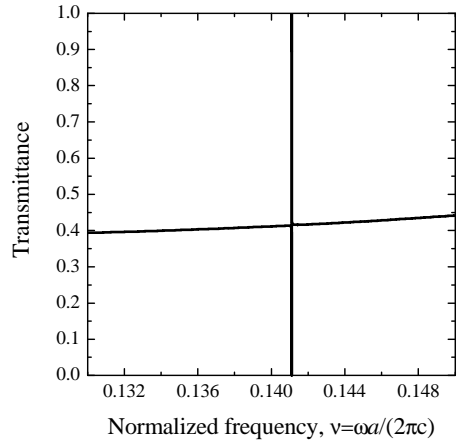

Fig. 6. As in Fig. 2, but pertaining to the point-defected supercell in Fig. 1(c).

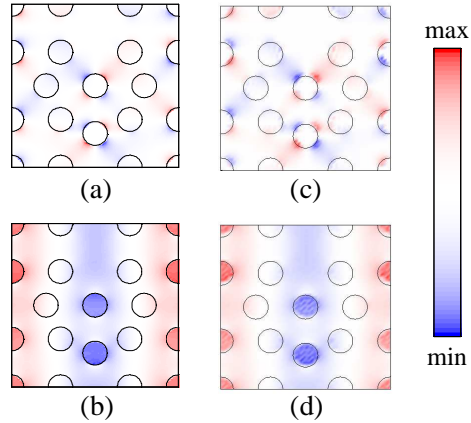

Fig. 7. As in Fig. 3, but pertaining to the point-defected supercell in Fig. 1(c). The resonant frequency is $\nu = 0.1411$, and the modal solution is labeled as c1 in Table 1.

A more interesting scenario comes up in connection with the dual configuration in Fig. 1(d), whose point-defect breaks the mirror symmetry along the *x*-axis, while preserving that along the *y*-axis. Via simple symmetry considerations, it is readily realized that the leaky modes supported by this structure can be obtained from those pertaining to that in Fig. 1(c) by simply interchanging the *x*- and *y*- axes. This considerably changes the relevant symmetry properties, and allows the excitation of *two* GRs, as shown in Fig. 8, which can be associated to two leaky modes in Table 1 (d2 and d3) that exhibit even symmetries along the *y*-axis (and are non-symmetric along the *x*-axis). The corresponding field maps are shown in Fig. 9. Once again, the modal solutions match very well the plane-wave response. It is interesting to note that, while one of the two uncoupled modes (d1) exhibits an odd symmetry along the *x*-axis (which clearly prevents its coupling), the other one (d4) exhibits an even symmetry along the *y*-axis and is non-symmetric along the *x*-axis, and thus may potentially couple. However, it was verified that the spatial distribution of its *y*-component is practically *zero-mean*, thereby yielding a negligible overlap with the impinging plane-wave.

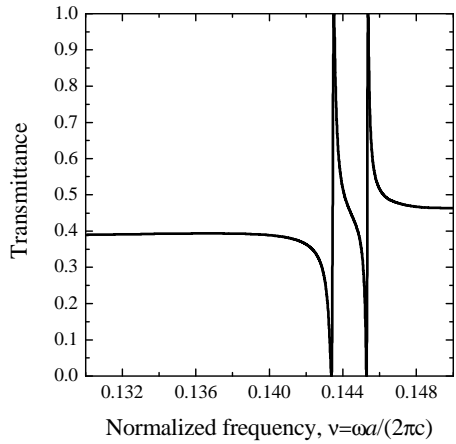

Fig. 8. As in Fig. 2, but pertaining to the point-defected supercell in Fig. 1(d).

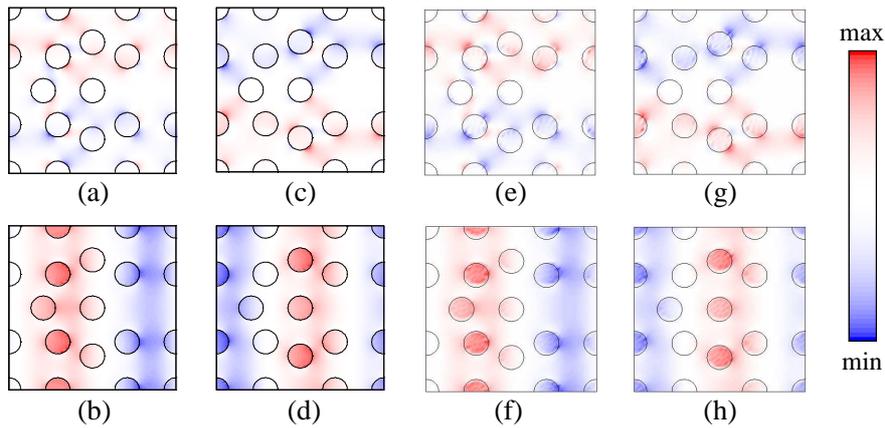

Fig. 9. Electric field maps (at *z*=0) corresponding to the two resonances in Fig. 8. (a), (b) $\nu = 0.1435$ (*x*- and *y*-component, respectively); (c), (d) $\nu = 0.1453$ (*x*- and *y*-component, respectively). (e)-(h) Corresponding modal solutions (labeled as d2 and d3, respectively in Table 1).

Finally, we consider the configuration in Fig. 1(e), featuring a point-defect that breaks the mirror symmetries along *both* the *x*- and *y*- axes. As expectable, and observable from Table 1, the four leaky modes supported by such structure are *non-symmetric* along both axes. As a result, they can all couple to the impinging plane-wave, giving rise to *four* resonances in the transmittance response, as shown in Fig. 10, with the corresponding field distributions shown in Figs. 11 (plane-wave-excited) and 12 (modal solutions). Note that, in view of their closeness, these resonances are not easy to resolve, and do not exhibit full amplitude excursions. Also in this case, a good match is found between the modal solutions and the plane-wave response, in terms of both resonant frequencies (cf. Table 1) and field distributions. Looking at the field maps in Figs. 11 and 12, the wealth of spatial distributions attainable is evident.

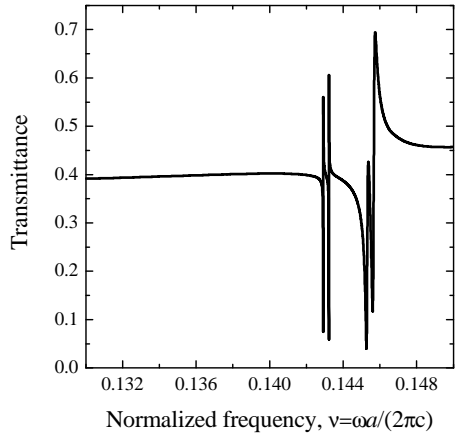

Fig. 10. As in Fig. 2, but pertaining to the point-defected supercell in Fig. 1(e).

*4.5 Remarks*

The above examples clearly illustrate the effects of breaking the supercell mirror symmetries in the excitation of otherwise uncoupled GRs, and the symmetry properties inherited by their spatial field distributions. In particular, for the assumed geometries and parameter configurations, breaking both the *x*- and *y*- symmetries allows the excitation of *all* the four GRs supported by the structure within the frequency range of interest. It is worth stressing that the focus of the above prototype studies was on the illustration of the basic phenomenologies, and no effort was made at this stage to optimize the parameter configurations.

To sum up, the aperiodically-ordered supercell considered in this paper, while not being the only conceivable or the "optimal" one, serves as a simple testbed to illustrate for the first time some intriguing GR tailor/control mechanisms based on the asymmetrization (via point defects) of the lattice geometry, instead of the hole shape. In particular, the above results clearly illustrate the anticipated extra degrees of freedom available by comparison with configurations based on point-defected *periodic* supercells. In this framework, one might wonder to what extent the above results hold when the supercell size is increased, and the structures accordingly tend to actual PQCs. Besides the underlying computational-affordability considerations, as well as the frequency restrictions imposed by the single-mode operation (cf. (1)), one would intuitively expect the effects of a *single* point-defect to become vanishingly small as the supercell size is increased. Therefore, extension of the above results to larger (PQC-like) structures could be more effectively pursued via the exploration of *composite* defect configurations.

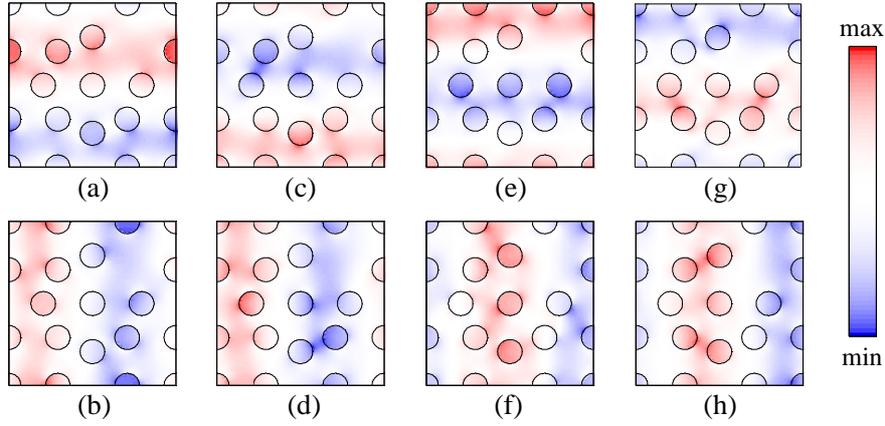

Fig. 11. Electric field maps (at $z=0$) corresponding to the four resonant frequencies in Fig. 10. (a), (b) $\nu = 0.1429$ ($x$- and $y$- component, respectively). (c), (d) $\nu = 0.1432$ ($x$- and $y$- component, respectively). (e), (f) $\nu = 0.1453$ ($x$- and $y$- component, respectively). (g), (h) $\nu = 0.1457$ ($x$- and $y$- component, respectively).

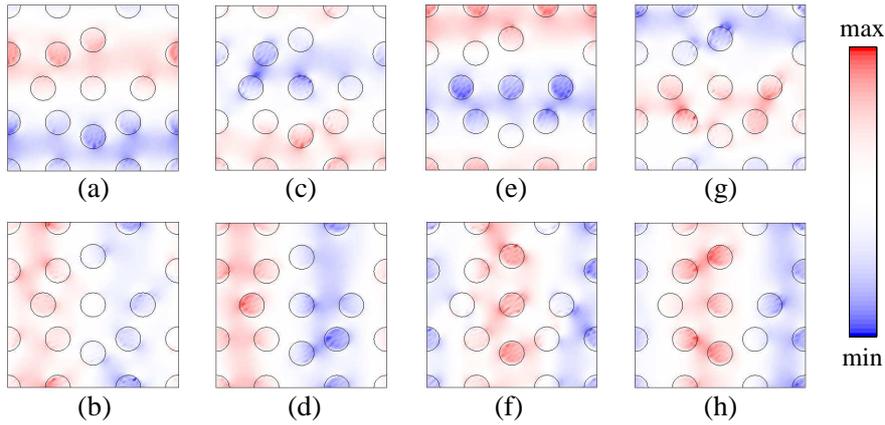

Fig. 12. Modal solutions (labeled as e1, e2, e3, and e4, respectively) corresponding to the plane-wave-excited fields in Fig. 11.

## 5. Conclusions and perspectives

In this paper, we have dealt with the investigation of GRs in plane-wave-excited PC slabs constructed by replicating point-defected aperiodically-ordered supercells based on the Ammann-Beenker (octagonal) tiling geometry. Via full-wave studies of the transmittance response and the modal structure, we have illustrated some representative effects induced by symmetry-breaking and symmetry-preserving point-defects, from which the *inherent richness* of the GRs excitable in such structures clearly emerges. The results of this prototype study indicate the possibility of controlling the GR coupling/uncoupling mechanisms, as well as the field distribution symmetries, by selectively breaking the supercell mirror symmetries via the judicious introduction of point-defects.

To the best of our knowledge, this represents the first example of application of such control mechanism, which appears as an attractive alternative (especially in terms of reduced sensitivity to fabrication tolerances) to those based on the asymmetrization of the hole shape [22], an may pave the way to new developments in the GR engineering. In this framework, current and future studies are aimed at exploring the technological viability of the proposed approach, via the parametric optimization of the candidate configurations, and their fabrication and experimental characterization. Extension of the above results in larger-supercell (PQC-like) structures could be effectively pursued via the exploration of *composite* defects.

**Acknowledgments**

This work was supported in part by the Campania Regional Government via a 2006 grant (L.R. N. 5 - 28.03.2002). The kind assistance of Prof. M. N. Armenise (Polytechnic of Bari, Italy) in the modal analysis is gratefully acknowledged.